\documentclass{pasj00}

\begin{document}
\SetRunningHead{M.~S.~Tashiro et al.}{Spectral Evolution in GRB Exponential Decays}
\Received{2011/05/09}
\Accepted{2011/10/07}

\title{Spectral Evolutions in Gamma-Ray Burst Exponential Decays Observed with Suzaku WAM}


%
\author{%
 Makoto~S. \textsc{Tashiro}\altaffilmark{1},
  Kaori  \textsc{Onda}\altaffilmark{1},
  Kazutaka \textsc{Yamaoka}\altaffilmark{2},
  Masahiro \textsc{Ohno}\altaffilmark{3},
  Satoshi \textsc{Sugita}\altaffilmark{4},
  Takeshi \textsc{Uehara}\altaffilmark{5},
  \and
  Hiromi \textsc{Seta}\altaffilmark{1}
}
\altaffiltext{1}{Department of Physics, Saitama University, 255 Shimo-Okubo, Sakura-ku, Saitama, 338-8570}
\email{tashiro@phy.saitama-u.ac.jp}
\altaffiltext{2}{Department of Mathematics and Physics, Aoyama Gakuin University, 5-10-1 Fuchinobe, Chuo-ku, Sagamihara, Kanagawa 252-5258} 
\altaffiltext{3}{JAXA/Institute of Space and Astronautical Science, 3-1-1 Yoshinodai, Sagamihara, Kanagawa 229-8510}
\altaffiltext{4}{EcoTopia Science Institute, Nagoya University, Furo-cho, Chikusa-ku, Nagoya, Aichi 464-8603}
\altaffiltext{5}{Department of Physics, Hiroshima University, 1-3-1 Kagami-Yama, Higashi-Hiroshima, Hiroshima, 739-8526}
 

\maketitle

\begin{abstract}
This paper presents a study on the spectral evolution of gamma-ray burst (GRB) prompt emissions observed with the Suzaku Wide-band All-sky Monitor (WAM). 
By making use of the WAM data archive, 6 bright GRBs exhibiting 7 well-separated fast-rise-exponential-decay (FRED) shaped light curves are presented and the evaluated exponential decay time constants of the  energy-resolved light curves from these FRED peak light curves are shown to indicate significant spectral evolution. 
The energy dependence of the time constants is well described with a power-law function $\tau(E) \propto E^\gamma$, where $\gamma \sim -(0.34\pm0.12)$ in average, although 5 FRED peaks show consistent value of $\gamma = -1/2$ which is expected in synchrotron or inverse-Compton cooling models.
In particular, 2 of the GRBs were located with accuracy sufficient to evaluate the time-resolved spectra with precise energy response matrices. 
Their behavior in spectral evolution suggests two different origins of emissions.
In the case of GRB~081224, the derived 1-s time-resolved spectra are well described by a blackbody radiation model with a power-law component.
The derived behavior of cooling is consistent with that expected from radiative cooling or expansion of the emission region. 
On the other hand, the other 1-s time-resolved spectra from GRB~100707A
is well described by a Band GRB model as well as with the thermal model. 
Although relative poor statistics prevent us to conclude, the energy dependence in decaying light curve is consistent with that expected in the former emission mechanism model.
\end{abstract}

\section{Introduction}\label{sec:intro} 
Gamma-ray bursts (GRBs) are the most energetic events known to occur in the universe. Their luminous yet variable behavior requires efficient radiation from relativistic outflows, although observations have not yet allowed us to obtain a clear understanding of the radiation process. Prompt emission spectra have been observed mostly in the gamma ray band and have been described with a power law, an exponentially cut-off power law, or a smoothly connected broken power law (GRBM)  (e.g., \cite{band93}; \cite{preece98}; \cite{preece00}; \cite{barraud03}; \cite{kaneko06}; \cite{kaneko08}; \cite{abdo09}; \cite{gonzalez09}). These non-thermal models are consistent with the energy spectral distributions from optically thin synchrotron or synchrotron self-Compton radiation produced by relativistic electrons accelerated through shock fronts in the outflows (\cite{tavani96}; \cite{cohen97}; \cite{schaefer98}; \cite{frontera00}; \cite{wang09}). 

\begin{table*}
\begin{center}
\caption{Event list} \label{tab:eventlist}
 \begin{tabular}{cccccc}
 \hline 
WAM trigger & GRB ID  & Trigger time (UT) & Position (J2000.0) & T90    & Reference\footnotemark[$*$]\\
 \hline 
\# 0058          & 050924       &  08:49:12              & not determined    & $8.41\pm0.05$ s &  (1) \\
\# 0408          & 060915       &  08:25:34              & not determined    & $43.15\pm0.07$ s  &  (2) \\
\# 0415          & 060922       &  17:21:21              & not determined    & $13.88\pm0.26$ s  & (3) \\
\# 0449          & 061101       &  21:26:28             & not determined     & $21.57\pm0.02$ s  & (4) \\
\# 1129          & 081224       &  21:17:55              & (\timeform{14h12m35.85s}, \timeform{+74d11m59.5s}) &$11.67\pm0.20$ s & (5) \\
\# 1555          & 100707A    &  00:46:36              & (\timeform{15h03m52.41s}, \timeform{-30d39m30.2s})  &$9.88\pm0.41$ s & (6)\\
 \hline 
\multicolumn{6}{@{}l@{}}{\hbox to 0pt{\parbox{170mm}{\footnotesize
\par\noindent
\footnotemark[$*$] (1)  Konus GRB \#2634; (2) Konus GRB \#2790; (3) Konus GRB \#2796 / SPI-ACS; (4) Konus GRB \#2808; (5) GCNC 8739; (6) GCNC 10950/10947
 }\hss}}  
 \end{tabular} 
\end{center}
\end{table*}

An energy peak seen in the cut-off power law or the broken power law is then attributed to the low energy end of the electron energy distribution. However, although non-thermal models have succeeded in describing the time-averaged spectra of prompt emissions, some observational results seem to be inconsistent with the models, such as in low-energy spectra exhibiting flatter slopes than those allowed in the optically thin synchrotron or synchrotron self-Compton (SSC) model obtained in some time-resolved spectral analyses (\cite{crider97}; \cite{preece02}; \cite{ghirlanda03}; \cite{gonzalez09}; \cite{ohno09}; \cite{onda10}). These findings have motivated researchers to study other emission mechanisms, such as small-pitch-angle synchrotron emission (\cite{lloyd00}; \cite{medvedev06}) and inverse-Compton (IC) scattering of a self-absorbed seed emission, although in these cases contradictions arise from the point of view of acceleration of electrons (\cite{spitkovsky08}; \cite{giannios09}), inefficiency of energy release (\cite{kobayashi97}; \cite{lazzati99}; \cite{nysewander09}) or the observed relation between luminosity and energy peaks (\cite{ramirez-ruiz02}).

Alternatively, an optically thick thermal (blackbody) component has been proposed in addition to the pure non-thermal (power-law) model (\cite{meszaros00}; \cite{meszaros02}; \cite{daigne02}; \cite{rees05}; \cite{ryde02}; \cite{ryde09}). The thermal component is expected to form a narrow energy peak in the sub-MeV range, while a broad non-thermal spectrum describes the wide-band emission of GRBs.
In \citet{ryde09}, the observed spectral time evolution of prompt emissions is described with a decrease in temperature of the thermal component in this thermal regime.

As previous papers have argued, spectral evolution provides a new perspective for understanding the physics in emission regions. 
The time scales of brightening and dimming are determined by several factors, such as the size and speed of the emission region and the acceleration/deceleration mechanism of electrons. The longest time scale among them dominates the observed light curve in the rise or decay phase, and its characteristic spectral evolution is observed in each phase. 
In case that the spectrum exhibits a simple power-law shape, no spectral evolution is expected by variation of the emission region size or beaming effect alone. 
Even with the ``curvature effect'', which accounts for the delayed photon emission from the high-latitude portion of the shell emission,
it requires only a self-similar temporal decay, such as $F_\nu \propto t^{-2-\alpha} \nu^{-\alpha}$ 
 (\cite{fenimore96}; \cite{kumar00}; \cite{demer04}; \cite{qin08}).
However, once a spectral break crosses the observation energy range by deceleration of the emission region, it is possible to create an apparent spectral evolution. In this regard, it has been shown in \citet{bbz09} that one can reproduce an exponentially decaying light curve with an apparent spectral evolution by using a model invoking a power-law spectrum with exponential cutoff.
On the other hand, heating/acceleration or cooling/deceleration of electrons naturally produce each of the characteristic spectral evolutions. The time scales are subject to the emission mechanism and the physical state in the emission region, such as temperature variation, and the time scale of shock acceleration, or radiation electron cooling. 

To reduce the number of parameters affecting the time evolution, here we focus on fast-rise-and-exponential-decay (FRED) like light curves, which are the most promising  for investigating the radiation process separately from the geometrical effects of the emission regions. The relatively fast rise implies that the time scale of geometrical variation is sufficiently short not to dominate the longer decaying process. The exponential decay time scale is thus expected to reflect the state evolution of the emission region, including the radiation cooling process. In this paper, we investigate luminosity-spectrum evolution by using data collected with the wide-band all-sky monitor (WAM; \cite{yamaoka09}) onboard Suzaku (\cite{mitsuda07}), and carried out spectral evolution study using energy-resolved light curves and time-resolved spectra.

\section{WAM observations and data selection}\label{sec:obs}

\begin{figure*}
  \begin{center}
    \FigureFile(165mm,165mm){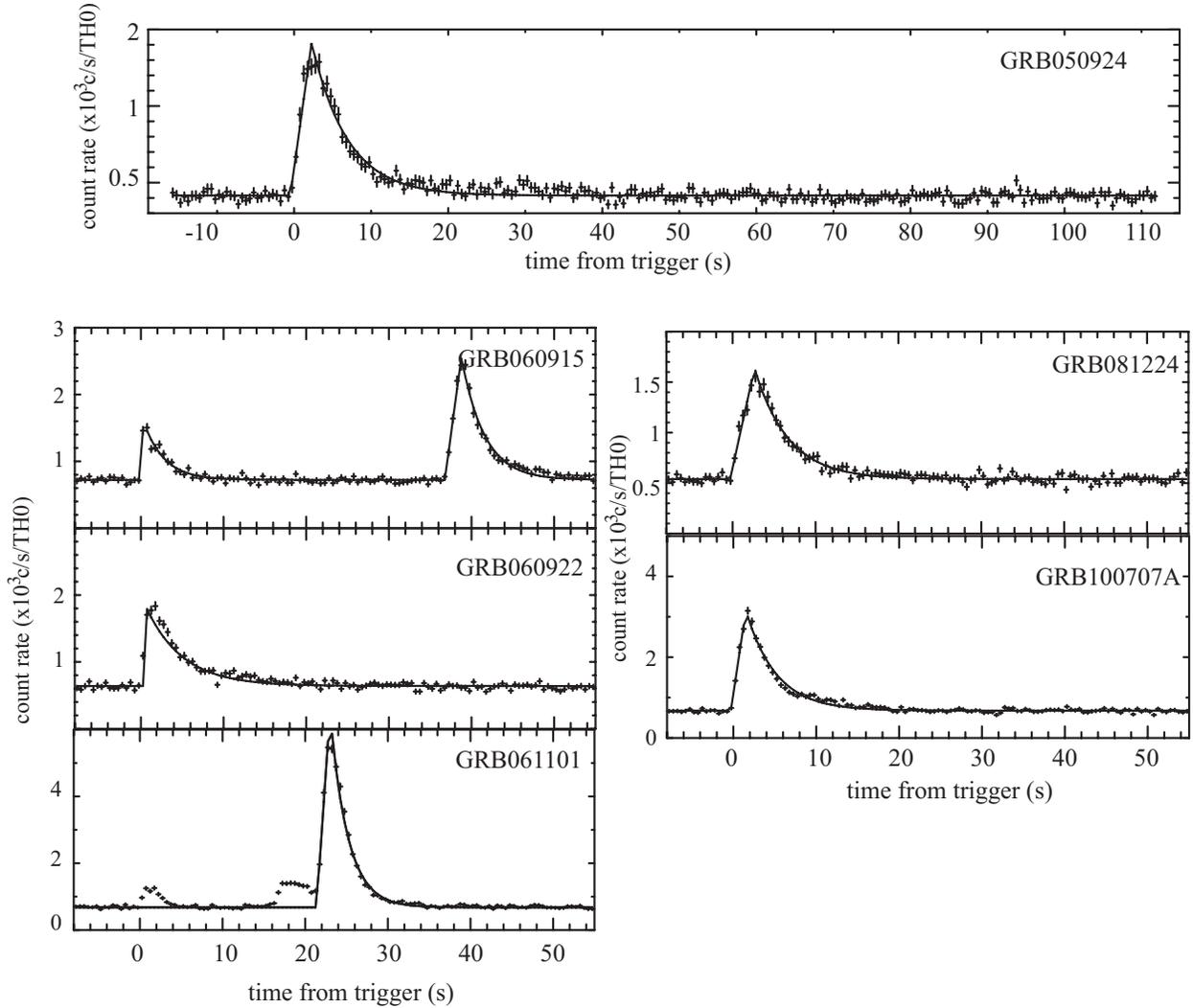}
  \end{center}
  \caption{TH0 light curves observed from the GRBs listed in table~\ref{tab:eventlist}. The light curves are evaluated with the FRED model described in \S~\ref{sec:lc}}. \label{fig:lc}
\end{figure*}

The Suzaku WAM is equipped with the largest effective area among all GRB detectors currently in orbit. 
The detection part of WAM consists of 20 BGO scintillator units surrounding four sides of the main well-type phoswitch counter units of the narrow field Hard X-ray Detector as their anti-coincidence shield counters.
The WAM monitors $2 \pi$ radian of sky in energies between 50 and 5000 keV.
The geometrical area of 800~cm$^2$ per one side and thick high-Z material ($Z_{eff} = 71$ for BGO) realized the effective area of 400~cm$^2$ even at the incident photon energy of 1~MeV.
The WAM triggers a GRB onboard by using the rising edge of the count rate by 5.5 rms of the previous 8~s background count rate, and stores 4 energy-resolved light curves for the period between 8~s (16~s before 20 March 2006) before and 56~s (112~s) after the trigger.
The WAM data are stored in two different formats, namely: ``TH'' data, which contains the 4 energy-resolved light curves with high time resolution of 1/64 (1/32)~s covering 64 ( 128)~s, and ``PH'' data.
The energy band of each TH light curve is labeled as TH0, TH1, TH2, or TH3, which nominally cover the 50--110 keV, 110--220 keV, 220--540 keV, and 540--5000 keV bands, respectively. 
``PH'' data with 54 energy bin spectra and a 1-s time resolution covering the energy band between 50 keV and 5 MeV, 
 which are continuously accumulated to provide background information for the Hard X-ray Detector. 
Since the WAM has no imaging capabilities, we can determine the angle of incidence 
 only for a GRB whose position is determined from simultaneous observation results from the interplanetary network (IPN) and/or Swift/BAT or Fermi/GBM. By using the position (and subsequently the incident angle), we generate a precise energy response for the spectral analysis with the PH data. 

We selected 6 GRBs with 7 isolated FRED-shaped pulses according to the following criteria in order to investigate the spectral evolution from all GRB events triggered onboard between August 2005 and December 2010.
(1) {\it Bright:} events whose TH2 count rate as observed by the most-irradiated side of WAM is higher than 1000 c~s$^{-1}$ at peak,
(2)  {\it Long decay:} events exhibiting 3-s time bins containing more than 100 c~s$^{-1}$ in energy bands up to TH2,
(3) {\it Fast-rise:} events exhibiting fast rise in less than 1/3 time interval of that of decay, in order to discriminate possible geometrical effect from the decay phase, and
(4) {\it Isolated:} events containing well-separated (without overlapping) peak(s) that are suitable for evaluating spectral evolutions from these light curves. 

In table~\ref{tab:eventlist}, information on selected events is summarized: GRB ID, trigger time, position (if determined), duration including 90\% of all observed photons (T90), and a reference to a simultaneous detection or a reported Gamma-ray burst Coordinate Network Circular (GCNC) number.
The T90 values are taken from the table published by the WAM team\footnote{http://www.astro.isas.ac.jp/suzaku/research/HXD-WAM/WAM-GRB/}.
Since GRB~081224 was observed with Fermi-GBM (\cite{GCN8723}) and with Swift-XRT (\cite{GCN8725}) in the afterglow phase, the position was well determined. GRB~100707A was observed with Knous-Wind \citep{GCN10948}, Fermi-GBM/LAT (\cite{GCN10944};\cite{GCN10945}) and the position was sufficiently determined for the purpose of this study with the IPN \citep{GCN10947}. 
Only for the both two events, we were able to generate energy responses using the positions as determined. A spectral analysis for each of these objects is presented in section~\ref{sec:spec}.
We used XSPEC version 12.5.1 for spectral analysis \citep{arnaud96}. All quoted errors are the 90\% confidence level.

\begin{table*}
 \caption{Decay power-law index for each TH band} \label{tab:beta}
\begin{center}
 \begin{tabular}{cccc}
\hline \hline 
GRB ID  & $\beta_0$     & $\beta_1$      & $\beta_2$            \\
         & ($\chi^2/$d.o.f.) & ($\chi^2/$d.o.f.) & ($\chi^2/$d.o.f.)   \\
 \hline 
050924  & $-0.82\pm0.05$ & $-0.91\pm0.05$ & $-0.86\pm0.10$ \\
         & $(71.5/ 16)$   & $(58.9/16)$      & $(40.5/ 12)$    \\
060915(1) & $-0.87\pm0.15$ & $-1.38\pm0.18$ & $-0.88\pm0.24$ \\
          & $(34.7/8 ) $   & $(21.9 / 6)$      & $(3.1 / 6)$    \\
060915(2)  & $-1.06\pm0.07$ & $-1.03\pm0.18$ & $-1.30\pm0.63$ \\
          & $(26.8 / 18)$   & $(4.14 / 6)$      & $(0.14 / 2 )$ \\
060922   & $-0.94\pm0.05$ & $-1.22\pm0.05$ & $-1.39\pm0.15$ \\
          & $(66.0 / 16 )$ & $(93.2 / 14)$      & $(9.80 / 6 )$ \\
061101   & $-1.49\pm0.04$ & $-1.38\pm0.11$ & $-1.85\pm0.62$ \\
          & $(171 / 12)$  & $(13.6 / 6 )$      & $(8.66/2)$\\
081224   & $-0.87\pm0.06$ & $-1.06\pm0.07$ & $-1.12\pm0.15$ \\
          & $(30.1 /14)$   & $(36.6 / 11)$    & $(8.99 / 6 )$ \\
100707A  & $-0.84\pm005$ & $-0.97\pm0.08$ & $-1.33\pm0.27$ \\
          & $(45.6 / 9 )$  & $(3.01 / 6)$   & $(9.3 / 6)$\\
 \hline 
 \multicolumn{4}{@{}l@{}}{\hbox to 0pt{\parbox{170mm}{\footnotesize
The derived time constant $\beta_n$ ($f(t) \propto t^{\beta}$) is shown for each TH$n$ band \\(see text).}\hss}} 
\end{tabular} 
\end{center}
\end{table*}

\section{Light curves}\label{sec:lc} 

In this section, we examine spectral evolution of the selected 7 FRED-shaped pulses using energy-resolved light curve (TH data).
Figure~\ref{fig:lc} shows all the TH0 light curves from the selected events, including the background. 
Here we employed light curves binned to 0.5-s resolution in order to evaluate possible trend of spectral variability with sufficient statistics.
We evaluated the averaged background count rate ($C$) by fitting each light curve ($f(t)$) in TH0 -- 3 separately, as shown in figure~\ref{fig:lc} for TH0, with the following function of $t$:
\begin{displaymath}
f(t) = \left\{ \begin{array}{ll}
		C                                                                                   &  (t  < t_{\rm s} ) \\
		\frac{t - t_{\rm s}}{t_{\rm p} - t_{\rm s}} P_0 + C & (t_{\rm s} < t < t_{\rm p}) \\
		P_0 e^{-\frac{t-t_{\rm p}}{\tau}} + C                      & (t_{\rm p} < t),\\
		\end{array}
	\right.
\end{displaymath}
where $t_{\rm s}$ and $t_{\rm p}$ are the start time of the rising and the peak time; further, $P_0$ and $\tau$ are the peak count rate and the decay time constant, respectively.
Since the light curves of GRB~061101 exhibit some activity before the FRED like peak, we omitted the time region before the peak from the fitting procedure.
These evaluation with all the parameters left free shows that most light curves rise within 1 or 2~s but do not provide satisfactory statistics to discuss energy dependence.

\begin{figure*}
  \begin{center}
    \FigureFile(70mm,70mm){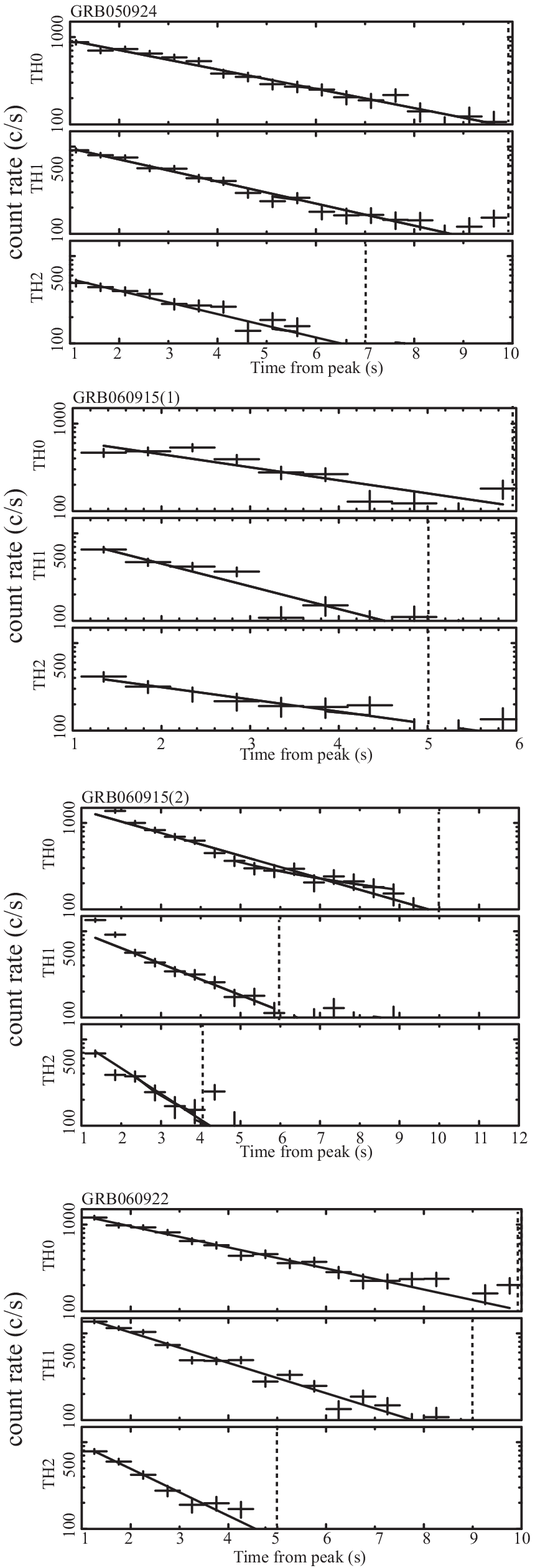}
    \FigureFile(70mm,50mm){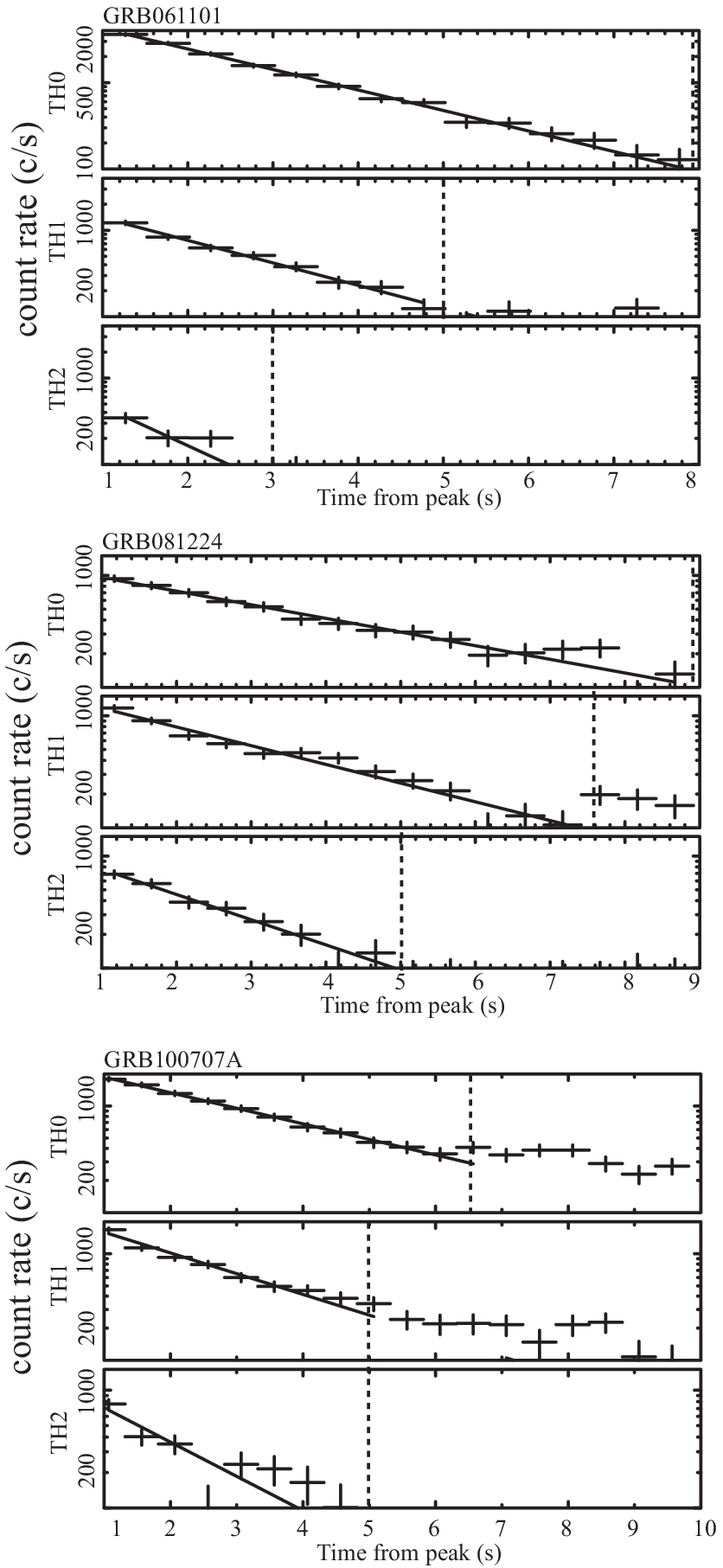}
  \end{center}
  \caption{FRED exponential decay features of the selected GRBs. The background subtracted light curves binned to 0.5-s resolution for TH0, TH1, and TH2 bands are shown in each top, middle and bottom panels. The best-fit results are shown as solid lines and the end of evaluated time regions are indicated in dashed lines.}\label{fig:fred}
\end{figure*}

Then, we concentrated on the decaying portions after we subtracted the background count rate evaluated above.
We omitted both time bin including the $t_{\rm p}$ evaluated for TH0, which exhibits the latest time in each pulse, and the next time bin  to avoid the possible effect by the rise portion.
The end of the time regions were limited by considering statistical and systematical uncertainty.
Since the typical rms of the evaluated background countrate is $\sim 33$ count$~$s$^{-1}$, we omitted the time regions the lightcurve below 100 count$~$s$^{-1}$ to avoid possible background fluctuation.
Consequently, all TH3 light curves were omitted from the following analysis.

We firstly examined a {\it power-law} function fit for these background subtracted decay light curves.
We set the time origin at $t_{\rm p}$ in this power-law fit.
However, none of the selected light curves were represented by a power-law function, as we see in table~\ref{tab:beta} summarizing the best fit decay index and the chi-squared value.
Then we employed exponential decay functions to evaluate decaying time constant for each TH lightcurve. 
We left the paramter $P_0$ free here.
The exponential decay model succeeded in fitting all of the energy-divided light curves as shown in figure~\ref{fig:lc}.
Note we carefully ignored small ($\sim 200$ c~s$^{-1}$ in each band) post FRED activity in GRB~100707A to obtain successful fit in the estimation of time constant.
Table~\ref{tab:tau} summarizes the best fit values and chi-squared values.

Interestingly, as we see in table~\ref{tab:tau} and figure~\ref{fig:fred}, all events exhibit shorter time scales in the higher energy bands. 
In order to evaluate this trend quantitatively, we fitted the derived time constants of decays with a power-law function ($\tau = N E^\gamma$) and summarized the obtained best-fit value of $\gamma$ within the 90~\% confidence level in table~\ref{tab:tau}.
The averaged value and its 90~\% confidence level of $\gamma$ is $-(0.34\pm0.12)$ ($\chi^2/$d.o.f.$=3.7/6$). 
It is noteworthy that the values of $\gamma$ as determined for 5 of the 7 FRED peaks are consistent with the average value of $-1/2$, which is the value expected for synchrotron or inverse-Compton cooling, while the rest 2 FRED peaks rejected.

\begin{table*}
 \caption{Exponential decay time constants for each TH band and the energy index} \label{tab:tau}
\begin{center}
 \begin{tabular}{cccc|c}
\hline \hline 
GRB ID  & $\tau_0$ (s)	& $\tau_1$ (s) & $\tau_2$ (s)             & $\gamma$\footnotemark[$\dagger$]    \\
         & ($\chi^2/$d.o.f.) & ($\chi^2/$d.o.f.) & ($\chi^2/$d.o.f.)   & ($\chi^2$) (d.o.f. = 1) \\
 \hline 
050924  & $3.93\pm0.31$ & $3.43\pm0.25$ & $3.19\pm0.44$  & $-0.14_{-0.16}^{+0.13}$ \\
         & $(13.4/16)$   & $(16.3/16)$      & $(18.7 / 12)$              & (0.07)\\
060915(1) & $2.92\pm0.53$ & $1.68\pm0.25$ & $3.12\pm1.01$ &$-0.48_{-0.51}^{+0.47}$ \\
          & $(21.5 /8 ) $   & $(14.7 / 6)$      & $(3.62 / 6)$               & (3.99)\\
060915(2)  & $3.30\pm0.31$ & $2.39\pm0.46$ & $1.46\pm0.92$ & $-0.45_{-0.43}^{+0.44}$\\
          & $(23.4 / 18)$   & $(1.69 / 6)$      & $(0.47 / 2 )$              & (0.06)\\
060922   & $3.58\pm0.31$ & $2.48\pm0.15$ & $1.60\pm0.21$  & $-0.50_{-0.14}^{+0.46}$ \\
          & $(27.1 / 16 )$ & $(29.9 / 14)$      & $(5.05 / 6 )$            & (0.16)\\
061101   & $1.83\pm0.07$ & $1.67\pm0.15$ & $0.96\pm0.35$   & $-0.22_{-0.18}^{+0.16}$\\
          & $(7.25 / 12)$  & $(3.20 / 6 )$      & $(7.13/2)$                & (1.60) \\
081224   & $3.56\pm0.32$ & $2.60\pm0.21$ & $1.93\pm0.31$ & $-0.40\pm0.18$\\
          & $(10.9 /14)$   & $(16.9 / 11)$    & $(2.93 / 6 )$               & (0.004)\\
100707A  & $2.97\pm0.20$ & $2.23\pm0.22$ & $1.49\pm0.40$  & $-0.40_{-0.21}^{+0.19}$\\
          & $(1.41 / 9 )$  & $(10.1 / 6)$   & $(16.0 / 6)$                   & (0.13)\\
 \hline 
 \multicolumn{5}{@{}l@{}}{\hbox to 0pt{\parbox{170mm}{\footnotesize
The derived time constant $\tau_n$ ($f(t) \propto e^{-t/\tau_n}$) is shown for each TH$n$ band (see text).
\par\noindent
\footnotemark[$\dagger$] The energy index of power-law function ($\tau \propto E^{\gamma}$) to evaluate the time constant $\tau_n$. \\
Note that the degree of freedom is unity since the evaluation is carried out for the \\
three energy divided time constants (see text).}\hss}} 
\end{tabular} 
   \end{center}
\end{table*}

\begin{figure*}

  \begin{center}
    \FigureFile(100mm,100mm){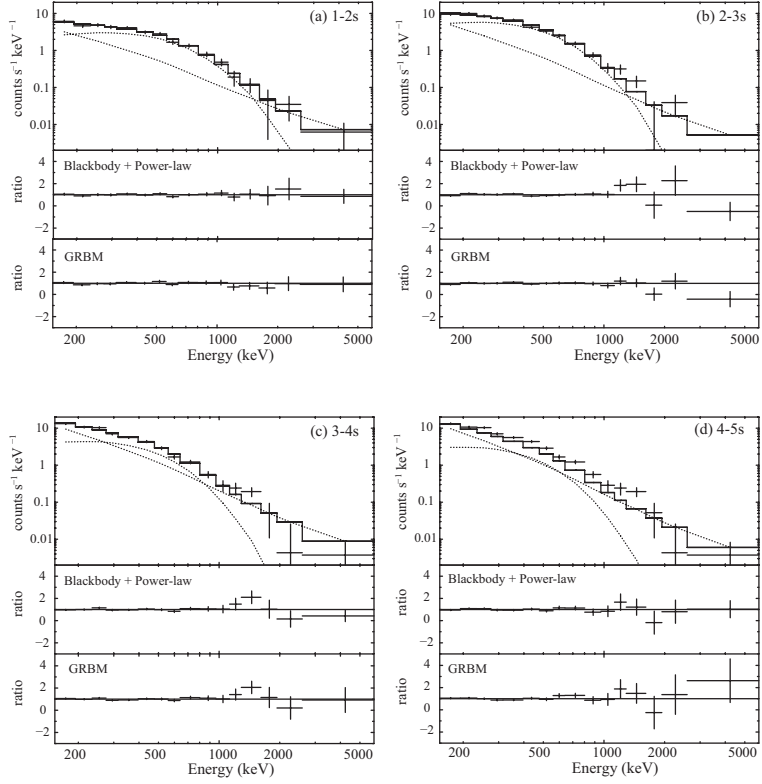}
  \end{center}
  \caption{The time-resolved folded spectra of GRB~081224 in the four second decay phases. Each 1-s integrated spectrum is fitted with a power law combined with a blackbody radiation model. The dotted lines represent the two components, while the solid histogram represents the sum of the two components. Ratio to the best fit models of BBPL and GRBM are shown in each middle and lower panels, respectively.}\label{fig:spec081224}
\end{figure*}

\section{Time-resolved spectrum}\label{sec:spec} 
As mentioned in section~\ref{sec:obs}, the positions of GRB~081224 and GRB~100707A were determined with an accuracy sufficient to generate reliable energy response matrices. 
In this section, we analyze the respective time-resolved spectra to examine the spectral behavior in detail.
Exposure time of each spectrum is 1-s which is the time resolution of PH data of WAM.
The time region here we present is limited to the time region that we showed in TH2 light curves in figure~\ref{fig:fred}.
In order to avoid uncertainty in generating the energy response, we ignored the bins below 150~keV in the following spectral analysis \citep{yamaoka09}, while we used the upper most energy bin covering up to 500 to 520 keV depending on the detector gain at each observation as provided by the WAM calibration team.

Figure~\ref{fig:spec081224} shows four 1-s time-resolved spectra from GRB~081224 in the decay phase.
We fit the four time-resolved spectra simultaneously to evaluate time trend of spectral parameters.
Here we examined the spectra with a single power-law model (PL), an exponential cut-off power-law model (CUT), GRBM (section 1) or a blackbody radiation with a power-law-shaped non-thermal emission model (BBPL).
While PL failed to describe the spectra with unacceptable chi-squared value of $\chi^2$/d.o.f. $=527.9/59$, CUT gave marginal fit with $68.6/55$.
On the other hand, the either of other two models --- GRBM and BBPL provided significant improvement in fitting the data as are shown in figure~\ref{fig:spec081224}.

\begin{figure}
  \begin{center}
    \FigureFile(60mm,60mm){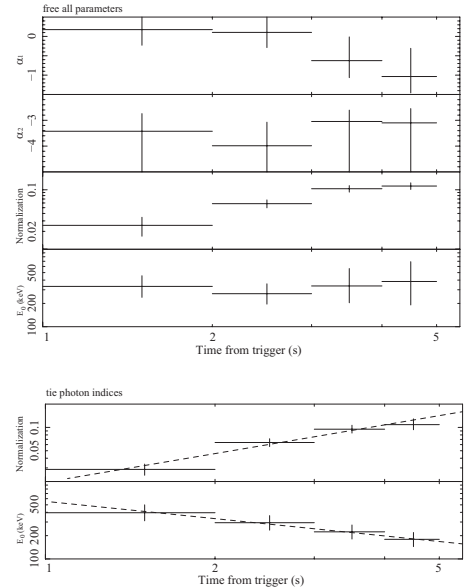}
  \end{center}
  \caption{The time history of derived best-fit parameters from the time-resolved spectra of GRB~081224 as fitted with GRBM. The upper four panels show the results at four free parameter fit, while lower two panels show the derived best fit parameters of normalization and the characteristic break energy ($E_0$) for each second. In the latter case, the photon indices ($\alpha_1, \alpha_2$) below and above $E_0$ are tied for all the spectra (see text). The normalization constant at 100 keV  is in units of photons~s$^{-1}$cm$^{-2}$keV$^{-1}$ and $E_0$ is in keV, respectively. }\label{fig:param081224grbm}
\end{figure}

Followings are the notations we employed for the GRBM.
\begin{displaymath}
f(t) = \left\{ \begin{array}{ll}
		N_0 \left( \frac{E}{100  {\rm keV}} \right)^{\alpha_1} e^{-E/E_0}       &  ( E  < (\alpha_1 - \alpha_2) E)_0 ) \\
		N_0 \left( \frac{(\alpha_1 - \alpha_2) E_0}{ 100 {\rm keV}} \right)^{\alpha_1 - \alpha_2} & \\
                \times \left( \frac{E}{100 {\rm keV}} \right)^{\alpha_2} e^{-(\alpha_1 - \alpha_2)}
                                                                                                   & (E  > (\alpha_1 - \alpha_2)E_0)\\
		\end{array}
	\right.
\end{displaymath}
Figure~\ref{fig:param081224grbm} shows the time history of derived best-fit parameters.
We fitted the spectra in two steps.
First we left all parameters free at the fitting and obtained the upper four panels shown in figure~\ref{fig:param081224grbm}.
Although these parameters describe the spectra well ($\chi^2/$ d.o.f. $=42.3/48$), we cannot see any trends in the derived spectral parameters.
Then, in order to evaluate the spectral evolution quantitatively avoiding the spectral parameter coupling, we tied the two photon indices of the four spectra. The obtained best fit values of lower and higher energy band photon indices in the simultaneous fitting were $\alpha_1 = -(0.08^{+0.25}_{-0.30}) $ and $\alpha_2 = -(3.13^{+0.54}_{-0.29})$ with the chi-squared value of $\chi^2$/ d.o.f.$ = 58.7/ 54$.
The other parameters, left free for the four spectra, normalization at 100 keV ($N_0$) and the characteristic break energy ($E_0$) are shown in bottom two panels of figure~\ref{fig:param081224grbm}.
The derived time trend of each of two parameters is well represented by a power-law function of time, $N_0 \propto t^{+1.26\pm0.30}$ ($\chi^2$/d.o.f.$=1.34/2$) and $E_0 \propto t^{-(0.72^{+0.45}_{-0.51})}$ (0.06/2).
However, it is noticed that the trend of increasing of $N_0$  and the hint of high energy photon excess in the ratio shown in figure~\ref{fig:spec081224} (d).
The former suggests an artifact at describing ``soft excess'' with ``steepening power-law with increasing normalization'' or ``lowering $E_0$ with increasing normalization'' as we see in figure~\ref{fig:param081224grbm}. 
With the latter, suggesting high energy excess in the faintest phase, these features hint us a BBPL with fainting blackbody component describes the low and high energy excess better.

\begin{figure}
  \begin{center}
    \FigureFile(60mm,60mm){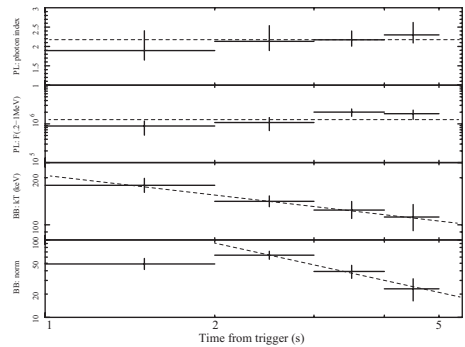}
  \end{center}
  \caption{The time history of derived best-fit parameters from the time-resolved spectra of GRB~081224 as fitted with a power law combined with a blackbody radiation model. From the top most to bottom panels show the time evolution of photon index, $0.2 - 1$MeV flux of the power-law component in unit of erg~s$^{-1}$cm$^{-2}$, $kT$ of the black body component in keV, and the normalization of the black body component in $L_{51}/D^2_{10}$ where $L_{51}$ is the source luminosity in units of $10^{51}$erg~s$^{-1}$ and $D_{10}$ is the luminosity distance to the source in units of 10 Gpc. The dashed line in each panel represents the best fit power-law function, whose best fit parameters are described in text.}\label{fig:param081224bbpl}
\end{figure}

\begin{table*}
 \caption{Time trends of best fit parameters for the time resolved spectra from GRB~081224} \label{tab:trend081224}
\begin{center}
 \begin{tabular}{ccc|ccc}
\hline \hline 
\multicolumn{3}{c|}{BBPL model}   & \multicolumn{3}{c}{GRBM}  \\
parameter  & $A$\footnotemark[$\dagger$] & $\eta$\footnotemark[$\dagger$]  &
parameter  & $A$\footnotemark[$\dagger$] & $\eta$\footnotemark[$\dagger$] \\
 \hline 
$\alpha$ & $2.17^{+0.26}_{-0.28}$ &  0 (constant) &
$\alpha_1$ & $-(0.08^{+0.25}_{-0.30}) $   & 0  \\

$F(0.2-1{\rm MeV})$ erg~s$^{-1}$cm$^{-2}$ & $(1.3\pm0.3) \times 10^{-6}$ & 0  & 
$\alpha_2$  & $-(3.13^{+0.54}_{-0.29})$ &  0 \\

$kT$ keV  & $ 210^{+60}_{-50}$ & $-(0.43^{+0.27}_{-0.28})$ &
$E_0$ keV & $550^{+330}_{-260}$ & $-(0.72^{+0.45}_{-0.51})$ \\

$L_{51}/D^2_{10}$\footnotemark[$\ddagger$] & $70^{+25}_{-21}$& $-(1.60^{+0.90}_{-0.75})$ &
$N_0$ & $0.019\pm0.006$ & $+1.26\pm0.30$ \\
\hline 
\multicolumn{6}{@{}l@{}}{\hbox to 0pt{\parbox{170mm}{\footnotesize
\par\noindent
\footnotemark[$\dagger$] The time trend of each parameter are described in a power-law function $A t^{\eta}$.\\
\footnotemark[$\ddagger$]  $L_{51}/D^2_{10}$ where $L_{51}$ is the source luminosity in units of $10^{51}$erg~s$^{-1}$ and $D_{10}$ is the luminosity distance to the source in units of 10 Gpc. See text for the other paramters.}\hss}} 
\end{tabular} 
\end{center}
\end{table*}

\begin{figure*}
  \begin{center}
    \FigureFile(100mm,100mm){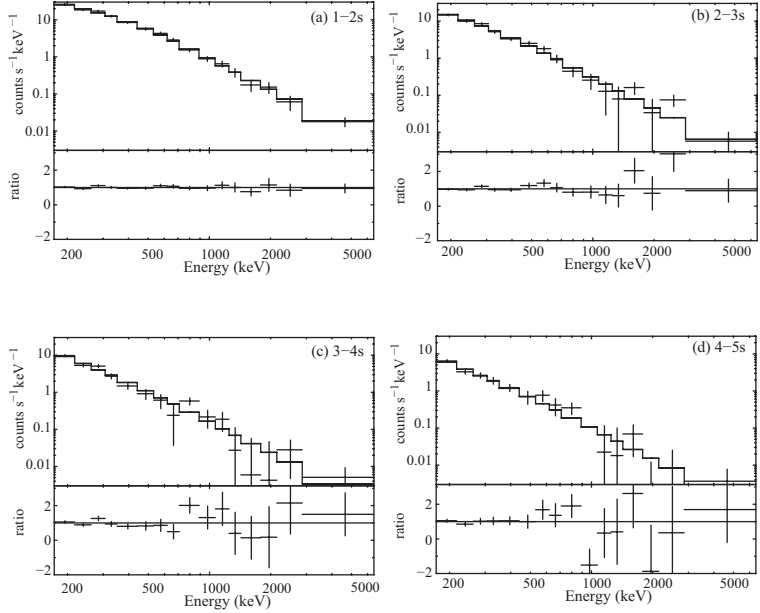}
  \end{center}
  \caption{The time-resolved folded spectra of GRB~100707A in each four second of the decay phase. Each 1-s integrated spectrum is fitted with a GRBM, which is shown in solid histograms. Ratios to the best fit model are shown in each lower panel.}\label{fig:spec100707}
\end{figure*}

The BBPL succeeded to describe the time resolved spectra with $\chi^2 /$ d.o.f. $ = 56.0 / 48$.
The parameters of the derived best fit with BBPL are shown in figure~\ref{fig:param081224bbpl}, which implies a relatively constant power-law component with a gradually cooling blackbody component. The photon indices of the derived best fit are consistent with a constant value of $2.19^{+0.16}_{-0.19}$ ($\chi^2/$d.o.f.$ = 1.16/ 4$), while the 200--1000 keV flux shows an extremely small variation around the average value of $(1.22\pm0.29)\times10^{-6}$ erg~s$^{-1}$~cm$^{-2}$ ($\chi^2/$d.o.f.$= 10.2/4$). 
Interestingly, the derived behavior of $kT$ is well reproduced by a power-law function of time with an index of $-(0.43^{+0.27}_{-0.28})$ ($\chi^2/$d.o.f.$ = 0.02/2$), whereas in the blackbody flux a monotonous power-law decay in the last 3 s with an index of $-(0.47\pm0.33$ ($\chi^2/$d.o.f.$= 8.3/2$) although we see obvious deviation at the first bin.
All derived time trend of the spectral parameters are summarized in table~\ref{tab:trend081224}.
Here, the results indicate that the energy dependence of the decay time constants in the energy-resolved light curves should be attributed to the cooling of the blackbody component qualitatively; however, from a quantitative point of view, the energy dependence is obscured by a fairly constant non-thermal component.

\begin{figure}
  \begin{center}
    \FigureFile(60mm,60mm){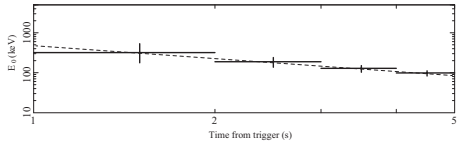}
  \end{center}
  \caption{The time history of derived best-fit break energies ($E_0$) in keV for the time-resolved spectra of GRB~100707A. The other three spectral parameters ($\alpha_1, \alpha_2$, and the normalization constant) are tied. The dotted line indicates the best fit power-law function (see text).}\label{fig:param100707}
\end{figure}

The four 1-s time-resolved spectra from the decay phase in GRB~100707A exhibit no significant preference for the four and two parameter fit of GRBM ($\chi^2/$d.o.f.$= 62.4/48$ and $=62.8/54$, respectively) and BBPL ($\chi^2/$d.o.f.$=60.6/48$). 
However we cannot find any time trends with four nor two free parameters for each spectrum fit, because of the large uncertainty of the best fit parameters. Moreover, similar chi-squared values for different d.o.f. suggests reducing free parameters. 
We examined single parameter fits by choosing the turn-over energy $E_0$ for GRBM and tied the normalization, the first and second spectral index ($\alpha_1, \alpha_2$).
The time evolution of the best-fit value of  $E_0$ are summarized in figure~\ref{fig:param100707}. 
The parameter of $E_0$ exhibits a power-law-shaped decrease in time with an index of $-(1.1^{+0.6}_{-0.8})$  ($\chi^2/$d.o.f.$ = 0.03/2$) as summarized in figure~\ref{fig:param100707} and table~\ref{tab:trend100707}.   

\begin{table}
 \caption{Time trends of best fit parameters for the time resolved spectra from GRB~100707A} \label{tab:trend100707}
 \begin{tabular}{ccc}
\hline \hline 
\multicolumn{3}{c}{GRBM} \\
parameter  & $A$\footnotemark[$\dagger$] & $\eta$\footnotemark[$\dagger$] 
\\
 \hline 
$\alpha_1$ & $-(0.83^{+0.39}_{-0.57})$ & 0 \\
$\alpha_2$  & $-(2.52^{+0.12}_{-0.10})$ & 0  \\
$E_0$ keV & $500^{+510}_{-330}$ & $-(1.1\pm0.7)$\\
$N_0$ & $0.28\pm0.03$ & 0  \\
\hline 
\multicolumn{3}{@{}l@{}}{\hbox to 0pt{\parbox{170mm}{\footnotesize
\par\noindent
\footnotemark[$\dagger$] The time trend of each parameter are described in a power-law \\ function $A t^{\eta}$.
}\hss}} 
\end{tabular} 
\end{table}

\section{Discussion} 
In order to investigate the cooling/deceleration mechanism, we selected 7 well-separated asymmetric FRED-like peaks from 6 GRBs observed with WAM in accordance to the selection criteria described in section~\ref{sec:obs}.
As described in section~\ref{sec:lc}, all decays are described well with exponential decay function while they clearly rejected power-law decay.
We examined all of the energy-resolved light curves systematically and qualitatively with a linear-rise-exponential-decay function.
Next, we found the obtained decay time constants, which are well described with a power-law function of the energy.
All energy indices are negative, which implies shorter time constant at higher energy band.
Although as yet we cannot propose a clear interpretation for the light curves exhibiting $\gamma =0.34\pm0.12$ in average, we note that the 5 FRED peaks showing $\gamma = -1/2$ are consistent with the expected results for non-thermal synchrotron and/or inverse-Compton cooling \citep{rybicki}. 

On the other hand, as mentioned in section 1, no exponential decay can be obtained with a simple power-law spectrum produced in a decelerated region. Even if we consider the curvature effect for a cut-off power-law spectrum, it has been shown in \citet{bbz09} that the light curve exhibits a power-law type of decay in proportion to $t^{-(2+\alpha)}$ around the cutoff energy, where $\alpha$ denotes the power-law index of the source spectrum. Therefore, we conclude that neither the observed exponential decay nor its energy dependence as observed in the 7 selected GRBs are expected in the geometrical interpretation.
Therefore, at least for the cases selected under the abovementioned criteria, the decays are not dominated by geometrical effects.

Before concluding the emission mechanism or the time evolution of the physical parameters of the emission region, we investigated the time-spectral evolution from the two GRBs located with the aid of simultaneous observations.
The time-resolved spectra obtained from GRB~081224, exhibiting energy index of decay time constant $\gamma = -0.40\pm0.18$ in the light curves, are well reproduced with BBPL as well as GRBM.
However, it is not easy for us to draw a clear view of what makes both brightening with lowering characteristic break energy ($E_0$).
 
The observed exponential decay in the energy-resolved light curves can also be attributed to the time evolution of the thermal component, however, the energy dependence observed in the light curve was under estimated by the relatively constant non-thermal component.
It is also noteworthy that the derived temperature of the blackbody radiation component decreases in proportion to $t^{-(0.43^{+0.27}_{-0.28})}$ (table~\ref{tab:trend081224}).
In the case of blackbody radiation cooling, the temperature variation at the source rest frame is governed by the following equation. Here, we assume conservation of the number of electrons in the emission region, and thus the product of the number density of electrons and the volume of the emission region $nV$ is constant.
\begin{eqnarray*}
	\frac{dE}{dt} &=& - nVk \frac{dT}{dt} = - \sigma S T^4 \\
	T &=&  \left(\frac{3 \sigma S}{nVk} \right)^{-1/3} t^{-1/3},
\end{eqnarray*}
where $E$, $t$, $T$, and $S$ denote the total thermal energy, time, temperature, and emission area, respectively.
We naturally regard the temperature decay (e.g $T \propto t^{-1/3}$) observed from the spectral variation in GRB~081224 as a result
of blackbody radiation cooling and/or a increase in the emission area with time (e.g. $S \propto t^2$).

On the other hand, the time-resolved spectra obtained from the other GRB~100707A are well reproduced with GRBM, although they do not accept a thermal model.
The photon indices of the lower and higher energy bands ($\alpha_1, \alpha_2$) are consistent with those expected from the standard synchrotron regime.
The decay index of $E_0$ is also consistent, though the uncertainty is large, with the distribution reported in \citet{peng09}, where FRED samples are examined using the data of the Burst and Transient Source Experiment on the Compton Gamma Ray Observatory ((decay index of $E_0$) $\sim -0.73$ or $-0.76$ with a standard deviation of 0.22).
The light curve indicating $\gamma =-(0.40^{+0.21}_{-0.19})$ suggests 
a non-thermal synchrotron or inverse-Compton process with a cooling time scale of $\tau=E^{-1/2}$  \citep{rybicki}.
Adopting the synchrotron and/or inverse-Compton cooling regime, we estimated the magnetic field strength ($B$) as a function of the cooling frequency ($\nu_{\rm c,obs}$) and the cooling time scale ($t_{\rm c,obs}$) in the observer frame. 
By employing the source redshift ($z$), the Doppler factor ($\delta \equiv \Gamma/(1-\beta_c \cos\theta)$, where the Lorentz factor, velocity in the unit of light speed, and the angle between the jet and line of sight are denoted as $\Gamma$, $\beta_c$, and $\theta$, respectively), and the Compton parameter $Y$, $B$ is expressed as a modification of Eq. (2) presented in \citet{KM08}, 
\begin{displaymath}
B = \left( \frac{ 18 \pi q m_{\rm e} c (1+z) }{  \sigma_{\rm T}^2 \nu_{\rm c,obs} t_{\rm c, obs}^2 \delta  (1 + Y)^2 }
     \right)^{1/3},
\end{displaymath}
where  the physical constants electron charge, electron mass, speed of light, and Thomson cross-section are denoted as  $q$, $m_{\rm e}$, $c$, and $\sigma_{\rm T}$, respectively. 
This equation can be rewritten as, 
\begin{eqnarray*}
B  &=& 1 \left( \frac{1+z}{2} \right)^{1/3} \left( \frac{\delta}{10^3} \right)^{-1/3} \left( \frac{1+Y}{2} \right)^{-2/3} \\
        && \times          \left( \frac{h \nu_{\rm c,obs}}{100 {\rm keV}}\right)^{-1/3} 
		\left( \frac{t_{\rm c,obs}}{3 {\rm s}} \right)^{-2/3} {\rm milli gauss,  or}\\
    &=& 20 \left( \frac{1+z}{2} \right)^{1/3} \left( \frac{\delta}{10^3} \right)^{-1/3} \left( \frac{1+Y}{10} \right)^{-2/3} \\
        && \times           \left( \frac{\nu_{\rm c,obs}}{10^{14} {\rm Hz}}\right)^{-1/3} 
		\left( \frac{t_{\rm c,obs}}{3 {\rm s}}\right)^{-2/3} {\rm milli gauss,}
\end{eqnarray*}
in cases where synchrotron or inverse-Compton emissions are observed in the observed gamma ray band, respectively.
Thus, a magnetic field with a respective strength of a few milligauss or a few tens milligauss can be expected for the two cases.
These values are significantly smaller than those estimated in previous works.
It is mainly because we employed relatively large cooling time scale events according to our selection criteria.
This may suggest that the GRBs we studied had small beaming factors ($\delta$).
Regarding the possible seed photons for the inverse-Compton process, it is necessary to consider not only synchrotron photons produced in the source (jet) rest frame, but also field photons outside the synchrotron electron field (e.g., \citet{tashiro09}).
In particular, it is noteworthy that the energy density of cosmic microwave background (CMB) photons, ($u_{\rm CMB}$), irradiating the GRB jet ($u_{\rm CMB} = 4.1 \times 10^{-13} (1+z)^4 \Gamma^3$, where $z$ and $\Gamma$ are the source redshift and the Lorentz factor of the jet, respectively) exceeds the energy density of the magnetic field.
Assuming $z=1$ and possible Lorentz factor e.g. $\Gamma = 100$ or 1000, we then obtain $u_{\rm CMB} = 6.5 \times 10^{-6}$ erg~cm$^{-3}$ or $= 6.5 \times 10^{-3}$ erg~cm$^{-3}$, while the energy density of the magnetic field becomes $u_{\rm B} \equiv B^2/(8\pi) = 4.0 \times 10^{-8}$ erg~cm$^{-3}$ or $u_{\rm B} = 4.0 \times 10^{-5} $ erg~cm$^{-3}$ for synchrotron gamma rays or inverse-Compton gamma rays, respectively.
Since the luminosity of synchrotron radiation and that of a CMB-boosted inverse-Compton process are proportional to $u_{\rm B}$ and $u_{\rm CMB}$, respectively, this requires that the luminosity of a CMB-boosted Compton process be much greater than that of synchrotron radiation in the extremely relativistic jets assumed for the GRB.
We need more precise information on the bulk Lorentz factor to conclude which the dominant seed photon source is.

The authors appreciate anonymous referee's very careful reading and many constructive suggestions.
The authors thank the Suzaku operations team and the WAM team for their effort work in making observations and performing calibrations. Part of this work was financially supported by the Ministry of Education, Culture, Sports, Science and Technology, Grant-in-Aid for Scientific Research No. 22340039.



\clearpage


\clearpage


\end{document}